\documentclass[english, preprint, amsmath,amssymb, aip]{revtex4-2}
\usepackage[T1]{fontenc}
\usepackage{float}
\usepackage{amsmath}
\usepackage{amssymb}
\usepackage{enumitem}
\usepackage{xcolor}
\usepackage{graphicx}
\usepackage{dcolumn}
\usepackage{bm}
\usepackage{hyperref}
\usepackage{setspace}
\usepackage{threeparttablex}
\usepackage{booktabs}
\usepackage{array}
\usepackage[title]{appendix}
     
\begin{document}

\title{Interplay between Static and Dynamic Disorder: Contrasting Effects on Dark State Population inside a Cavity}

\author{Robert F. Catuto}
\affiliation{Department of Chemistry and Biochemistry, 251 Nieuwland Science Hall, Notre Dame, Indiana 46556, United States}
\author{Hsing-Ta Chen}
\email{hchen25@nd.edu}
\affiliation{Department of Chemistry and Biochemistry, 251 Nieuwland Science Hall, Notre Dame, Indiana 46556, United States}

\date{\today}

\begin{abstract}
    Strong light-matter interactions between molecules and quantized electromagnetic fields inside an optical cavity open up novel possibilities, though inevitably influenced by disorder, an inherent attribute of realistic molecular systems. Here, we explore the steady-state optical response of molecular emitters within a lossy cavity, with a focus on the combined effects of static and dynamic disorder, as frequently observed in solution-phase experiments. By analyzing the transmission spectra, molecular energy change, and dark state population, we uncover the contrasting effects of static and dynamic disorder on the dark state population and its interplay with polariton states.
    We find that the Rabi splitting exhibits an inversion with increasing disorder strength where the maximum splitting is determined by the interplay of static and dynamic disorder.
    Furthermore, we identify a dark state-induced polariton linewidth narrowing, revealing a mechanism distinct from motional narrowing induced by frequency fluctuations. 
    These mechanistic insights highlight the critical role of dark states, establishing a foundation for future developments in the fields of polariton chemistry and strong coupling spectroscopy.
\end{abstract}
\maketitle


\section{Introduction}
Investigating the interplay between collective molecular optical response and local disorder effects has become a central theme in the study of strong coupling phenomena within organic molecular systems.\cite{khazanov_embrace_2023,george_controlling_2024,gera_effects_2022,allard_disorder-enhanced_2022,khan_analytic_2021} Strong light-matter coupling between a molecular transition possessing a dipole moment and the quantized electromagnetic field within an optical cavity leads to the formation of hybrid light-matter quasiparticles termed molecular polaritons.\cite{ribeiro_polariton_2018,xiang_molecular_2024,mandal_theoretical_2023} For an idealized molecular ensemble with uniform radiative coupling and identical transition frequencies, the collective optical response of molecular polaritons (such as superradiant emission and Rabi splitting) exhibits a characteristic dependence on molecular concentration or the number of coupled molecules ($N$).\cite{wright_versatile_2023,ahn_modification_2023,chen_exploring_2024} However, recent studies have shown that local disorder, including static inhomogeneity and dynamic frequency fluctuations, significantly affects strong coupling phenomena,\cite{houdre_vacuum-field_1996,celardo_interplay_2013,celardo_cooperative_2014,zhou_interplay_2023,chen_interplay_2022,chuang_anomalous_2024,liu_polaritons_2024} particularly within a solution phase.\cite{cohn_vibrational_2022,dutta_thermal_2024}

Solvent effects on molecules are often characterized by the frequency fluctuation correlation function (FFCF) within the context of two-dimensional infrared (2D-IR) spectroscopy.\cite{tokmakoff_vibrational_1995,shirley_experimental_2024} The slopes of the nodal lines (the center line of the ground state bleach peak) as a function of waiting time can be fitted to a single exponential decay function with a static offset $a_1 e^{-t/\tau_c}+a_0$.\cite{baiz_vibrational_2020,shirley_experimental_2024,pyles_revisiting_2024}
Here, $a_0$ and $a_1$ are unitless parameters characterizing the strength of the static and dynamic disorder, respectively, and $\tau_c$ is the correlation time of the frequency fluctuation.
Chuntonov \emph{et al.} systematically control $a_0$ by varying solvent mixtures and investigate molecular polaritons under varying static disorder strength.\cite{cohn_vibrational_2022}
Their results show an increase in Rabi splitting with static disorder strength in the small disorder regime, which can be explained by the mixing of the polariton states and optical dark states as induced by static disorder.\cite{zhou_interplay_2023}
However, at larger disorder strength, the Rabi splitting goes through an inversion, which cannot be captured using perturbation-based mechanism.\cite{cohn_vibrational_2022}
Furthermore, a counterintuitive narrowing of the polariton linewidth is observed in these experiments.

The key to understanding these intriguing phenomena lies in the investigation of optical dark states. Given the large manifold of dark states (approximately $N-1$) compared to the polaritons, their critical role in strong coupling phenomena is particularly significant in realistic systems.\cite{khazanov_embrace_2023,engelhardt_polariton_2023,pandya_microcavity-like_2021,liu_unlocking_2025,xiang_intermolecular_2020} The interactions between the polariton states and dark states result in polariton relaxation, provides potential catalytic pathways, and modifies polariton diffusion.\cite{gera_effects_2022,du_catalysis_2022,sharma_unraveling_2024,engelhardt_polariton_2023,ribeiro_multimode_2022,pandya_microcavity-like_2021} Recently, the interplay between the cavity photon lifetime and the molecular disorder has been investigated in terms of the stochastic lineshapes of polaritons\cite{climent_kubo-anderson_2024,yuen-zhou_linear_2024} 
That being said, in the presence of both static and dynamic disorder, the overall effect of the dark states on the Rabi splitting and polariton lineshape remains unclear.

In this paper, we investigate the effects of disorder on molecular emitters in an optical cavity, with a particular focus on the participation of dark states under the influence of both static and dynamic disorder.
First, we contrast the distinct dark state population for static and dynamic disorders, then discuss their combined effects on strong coupling characteristics, including Rabi splitting and polariton linewidth.
This paper is organized as follows. In Sec.~\ref{sec:model}, we formulate a model of a cavity-confined emitter ensemble with an energetic disorder and derive expressions for the cavity transmission and dark state population. In Sec.~\ref{sec:results}, we elucidate disorder effects in terms of the dark state population, focusing on Rabi splitting inversion. In Sec.~\ref{sec:interplay}, we clarify the combined effect of static and dynamic disorder on Rabi splitting and polariton linewidth. Finally, we conclude and discuss our findings in Sec.~\ref{sec:conclusion}.



\section{Model and Methods}\label{sec:model}

\subsection{Model Hamiltonian with Energetic Disorder}
Consider an ensemble of $N$ two-level emitters with the ground state $|g_j\rangle$ and the excited state $|x_j\rangle$ for $j=1\cdots N$ where the excited states can decay at a rate of $\Gamma_\text{loc}$. For the emitter ensemble, we denote the total ground state as $|G\rangle=\prod_{j=1}^N|g_j\rangle$ and the single excitation states as $|X_m\rangle=\prod_{j\neq m}^N|g_j\rangle\otimes|x_m\rangle$. The total ground state energy is $E_0$ and the single excitation energy is $E_m=E_0+\hbar\omega_m$.
This ensemble is placed inside an optical cavity supporting a single cavity photon mode of frequency $\omega_v$ and the cavity photon decay rate is $\Gamma_\text{cav}$.
We denote the vacuum state of the cavity as $|0_v\rangle$ and the single-photon state as $|1_v\rangle$. 
In the single-excitation manifold, we denote the dressed states by $|0\rangle=|G\rangle\otimes|0_v\rangle$, $|v\rangle=|G\rangle\otimes|1_v\rangle$, and $|m\rangle=|X_m\rangle\otimes|0_v\rangle$. 
The system Hamiltonian can be written as 
\begin{equation}
    \hat{H} = (E_v-i\Gamma_\text{cav})|v\rangle\langle v| + E_0|0\rangle\langle 0| + \sum_m (E_m-i\Gamma_\text{loc})|m\rangle\langle m|
    + \sum_m V_m(|v\rangle\langle m| + |m\rangle\langle v|) 
\end{equation}
Here $E_v=E_0+\hbar\omega_v$ and $V_m$ is the coupling strength between the cavity photon and the emitter excitation. 
For simplicity, we assume that the spatial size of the ensemble is smaller than the cavity photon wavelength (the long-wavelength approximation) so that all the emitters have identical coupling strength to the cavity photon mode, i.e. $V_m=g$ does not depend on $m$. 
We also adopt the rotating wave approximation, i.e. the coupling between emitters and cavity photon only includes energy-conserved terms, and higher-order excitations are neglected.

Here, the cavity photon mode is driven by an external continuous-wave (CW) field with the driving frequency $\omega$ and the amplitude $V_0$
\begin{equation}
    \hat{V}_\text{CW}(t) = V_{0} \cos(\omega t)(|0\rangle\langle v| + |v\rangle\langle 0|)
\end{equation}
Note that we assume that the emitter ensemble is not directly coupled to the external field (i.e. $|0\rangle$ and $|m\rangle$ are not coupled), and the external field is sufficiently weak so that the single photon excitation is valid.

In the absence of disorder, we set the emitters to be resonant with the cavity photon i.e. $E_m=E_v$ for all $m=1,\cdots,N$.
The cavity photon is coupled only to the symmetric emitter state $\sum_m|m\rangle$, which is considered bright to the cavity photon.  
Thus, the eigenvalues of the Hamiltonian lead to the upper and lower polaritons $E_\pm=E_v\pm\sqrt{N}g$, which corresponds to the superposition of the cavity photon and the bright state
\begin{equation}
    |P_\pm\rangle = \frac{1}{\sqrt{2}}|v\rangle \pm \sum_m \frac{1}{\sqrt{2N}}|m\rangle.
\end{equation}
In addition to the polariton states, there are $N-1$ degenerate eigenvalues $E_m=E_v$. The corresponding eigenstates are decoupled from the cavity photon, and are therefore considered dark with respect to the cavity photon mode. We denote the dark states as $|D_k\rangle=\sum_m\xi_{mk}|m\rangle$
where $\sum_m\xi_{mk}=0$ for $k=1,\cdots,N-1$.
It is important to note that, due to the coupling of the external field to the emitter mediated by the cavity photon, only the polariton states are populated; energetic disorder is a prerequisite for non-zero dark state population.

\subsection{Energetic disorder}
We consider the emitter's energetic disorder by introducing a time-dependent random frequency $\Omega_m(t)$ to the excitation energy
\begin{equation}
    \omega_m = \bar{\omega} + \Omega_m(t)
\end{equation}
where $\bar{\omega}$ is the average frequency corresponding to the transition between the ground and excited states of the emitters in the absence of disorder. The frequency fluctuation $\Omega_m(t)$ is a Gaussian stochastic random variable that follows the frequency fluctuation correlation function (FFCF)\cite{shirley_experimental_2024}
\begin{equation}
    \langle \Omega_m (t_1) \Omega_{m'}(t_2)\rangle = \delta_{mm'}(\sigma_{1}^2 e^{-|t_1-t_2|/\tau_c} +\sigma_0^2).
\end{equation}
where $\langle\cdots\rangle$ denotes the ensemble average.
Here static disorder is characterized by $\sigma_0$, which is associated with the variance of the emitter's energy distribution that does not depend on time.
The time-dependent energy fluctuation (dynamic disorder) is characterized by the disorder strength $\sigma_1$ and the correlation time $\tau_c$. Short correlation times imply rapid energy fluctuations of the emitter, while long correlation times imply nearly static disorder.
We ignore the cross-correlation by incorporating the Kronecker delta $\delta_{mm'}$.

To model both the static and dynamic disorder of the FFCF, we employ $\Omega_m (t)=A_m+B_m(t)$ where $A_m$ is a random variable following the Gaussian distribution with variance $\sigma_0^2$
\begin{equation}
    \text{Prob}[A_m]= \frac{1}{\sqrt{2\pi \sigma_0^2}}\exp{\left(-\frac{A_m^2}{2\sigma_0^2}\right)}
\end{equation}
$B_m(t)$ is chosen to be a Gaussian stochastic variable that satisfies $\langle B_m(t)\rangle=0$ and 
\begin{equation}
\langle B_m(t_1)B_m(t_2)\rangle=\sigma_1^2e^{-|t_1-t_2|/\tau_c}
\end{equation}
Note that, in the limit $\tau_c\rightarrow\infty$, the stochastic variable $B_m(t)$ is effectively a Gaussian distribution with $\langle B_m (t_1) B_{m}(t_2)\rangle \rightarrow\langle B_m^2\rangle = \sigma_1^2$.
In this case, we regard the overall FFCF follows an effective static disorder $\langle \Omega_m (t_1) \Omega_{m'}(t_2)\rangle \rightarrow\langle \Omega_m ^2\rangle=\sigma_1^2+\sigma_0^2=\sigma_\text{eff}^2$.

\subsection{Steady State Properties}
In terms of the dressed-state basis, the wavefunction for this single-excitation model can be represented as $|\psi(t)\rangle = c_0(t)|0\rangle + c_v(t)|v\rangle + \sum_m c_m(t)|m\rangle$. 
To simulate the steady-state properties, we choose the initial state to be $|\psi(0)\rangle = |0\rangle$ and propagate the Schr\"odinger equation $i\hbar\frac{d}{dt}|\psi(t)\rangle = [\hat{H}+\hat{V}_\text{CW}(t) ]|\psi(t)\rangle$ while keeping the ground state population constant (i.e. $|c_0(t)|^2 = 1$). 
At sufficiently long simulation time, we can calculate the following steady-state properties. 

The cavity transmission spectrum can be constructed from the outgoing flux from the cavity:
\begin{equation}
    J_\text{cav}(t) = \Gamma_\text{cav}\left\langle|c_{v}(t)|^2\right\rangle.
\end{equation}
For simplification, we assume that the cavity photon does not have a non-radiative decay channel. $\Gamma_\text{cav}$ is directly related to the cavity photon lifetime, which can be controlled by altering the quality factor of the cavity.
Similarly, the local relaxation flux can be calculated from the combined effect of the emitter's local relaxation:
\begin{equation}
    J_\text{loc}(t) = \Gamma_\text{loc}\left\langle\sum_{m}|c_{m}(t)|^2\right\rangle.
\end{equation}
The local decay rate $\Gamma_\text{loc}$ is corresponding to the homogeneous molecular bandwidth, which is the inverse of homogeneous dephasing of the emitter $T_2$.

The population of the dark states can be calculated by subtracting the projection of the upper and lower polariton states from the total wavefunction of the emitters:
\begin{align}
    P_\text{dark}(t) &= |\langle\psi'(t)|\psi'(t)\rangle|^2 -|\langle{P_+}|\psi'(t)\rangle|^2 - |\langle{P_-}|\psi'(t)\rangle|^2 \\
    &= \left\langle\sum_{m} |c_m(t)|^2 - \frac{1}{N}|\sum_{m}c_m(t)|^2\right\rangle
\end{align}
where $|\psi'(t)\rangle=\sum_m|X_m\rangle\langle X_m|\psi(t)\rangle$ is projected to the emitter part. 
We notice that, in contrast to the local relaxation flux, the dark state population is proportional to the variance of the coefficient $c_m$. Thus, a large dark state population implies a large variance of the wave vector coefficients. 

Lastly, we are interested in the emitter energy gain and loss as induced by the disorder effect. To obtain this, we first note that, in the absence of disorder, the emitter energy at the steady state is $\sum_{m} \hbar\bar{\omega}|c_m(t)|^2$, which is proportional to the local relaxation flux. In the presence of disorder, we can estimate the total energy difference by 
\begin{equation}
    E_\text{mol}(t) = \langle\psi'(t)|(\hat{H}-\hbar\bar{\omega})|\psi'(t)\rangle= \left\langle\sum_{m} \hbar\Omega_m(t)|c_m(t)|^2\right\rangle
\end{equation}
This steady-state total energy difference provides a measure of the emitter's energy gain and loss due to disorder effects.

\section{Disorder Effects in Terms of the Dark State Population}\label{sec:results}

\subsection{Model Parameters}

For all calculations in this paper, we consider $N = 40$ emitters placed in the cavity. 
The ground state energy is chosen to be $E_0= -1~\text{ps}^{-1}$ and the average excitation energy is set to be resonant with the cavity photon $\bar{\omega}=\omega_v= 1~\text{ps}^{-1}$.
We set the coupling strength between the excited state and the cavity photon mode to be $g={0.5}/{\sqrt{N}}~\text{ps}^{-1}$, which is normalized by $\sqrt{N}$ so that the Rabi splitting is $\Omega_R=1~\text{ps}^{-1}$ for the ordered case and not affected by the number of emitters.
We fix the external field intensity $V_0=5\times 10^{-4}~\text{ps}^{-1}$ and vary the driving frequency $\omega-\bar{\omega}$ in the range from $-1.5~\text{ps}^{-1}$ to $1.5~\text{ps}^{-1}$.

The decay rate of the local excited state is chosen to be $\Gamma_\text{loc} = 0.0083~\text{ps}^{-1}$, which comes from a representative vibrational lifetime ($T_2\sim120~\text{ps}$) of longer-lived vibrational modes, such as thiocyanate stretches in certain solvents.\cite{shirley_experimental_2024, wilderen_vibrational_2014}
The decay rate of the cavity photon is typically larger than the local relaxation rate $\Gamma_\text{cav} >\Gamma_\text{loc}$. We consider $\Gamma_\text{cav}$ to be in the range between $0.05~\text{ ps}^{-1}$ and $0.4~\text{ ps}^{-1}$, which already corresponds to an optical cavity with high quality factor, such as distributed Bragg reflector cavities.\cite{abbarchi_macroscopic_2013,xiang_manipulating_2019}

The disorder strength parameters ($\sigma_0$ and $\sigma_1$) are varied from $0$ to $0.5$, a range chosen to reflect the disorder strengths observed for methyl thiocyanate in various solvent environments.\cite{shirley_experimental_2024}
We explore the correlation time $\tau_c$ spanning the range from $10~\text{fs}$ to $10^3~\text{ps}$.
To generate Gaussian stochastic random variables $B_m(t)$, we use the Ornstein-Uhlenbeck process as outlined in Appendix A of Ref.~\onlinecite{zhou_interplay_2023}.
(\emph{i}) The initial value at $t_0$ is chosen according to a Gaussian distribution with standard deviation $\sigma_1$   
\begin{equation}\label{eq:B-gaussian}
        \text{Prob}[B_m(t_0)] = \frac{1}{\sqrt{2\pi \sigma_1^2}}\exp{\left(-\frac{B_m(t_0)^2}{2\sigma_1^2}\right)}
    \end{equation}
(\emph{ii}) The value at the next time step ($t_{n+1}$) is chosen according to the conditional probability:
    \begin{equation}
        \text{Prob}[B_m(t_{n+1})|B_m(t_n)] = \frac{1}{\sqrt{2\pi \sigma_1^2(1 - r_n^2)}}\exp{\left(-\frac{[B_m(t_{n+1}) - r_nB_m(t_n)]^2}{2 \sigma_1^2(1 - r_n^2)}\right)} 
    \end{equation}
where $r_n = \exp{\left(-(t_{n+1} - t_n)/\tau_c\right)} = \exp{(-dt/\tau_c)}$. Note that, when $\tau_c\rightarrow\infty$ i.e. $r_n\rightarrow1$, the conditional probability becomes a delta function $\text{Prob}[B_m(t_{n+1})|B_m(t_n)]\rightarrow\delta(B_m(t_{n+1})-B_m(t_n))$, thus $B_m(t_{n+1})\rightarrow B_m(t_n)$. 
Consequently, $B_m(t)$ becomes time-independent and follows Eq.~\eqref{eq:B-gaussian}, i.e. a static disorder with the disorder strength $\sigma_1$.
All steady-state properties were averaged over $300$ trajectories to ensure the results converged and well represented the ensemble.
  
    
    

\begin{figure}
    \centering
    \includegraphics[width=0.9\linewidth]{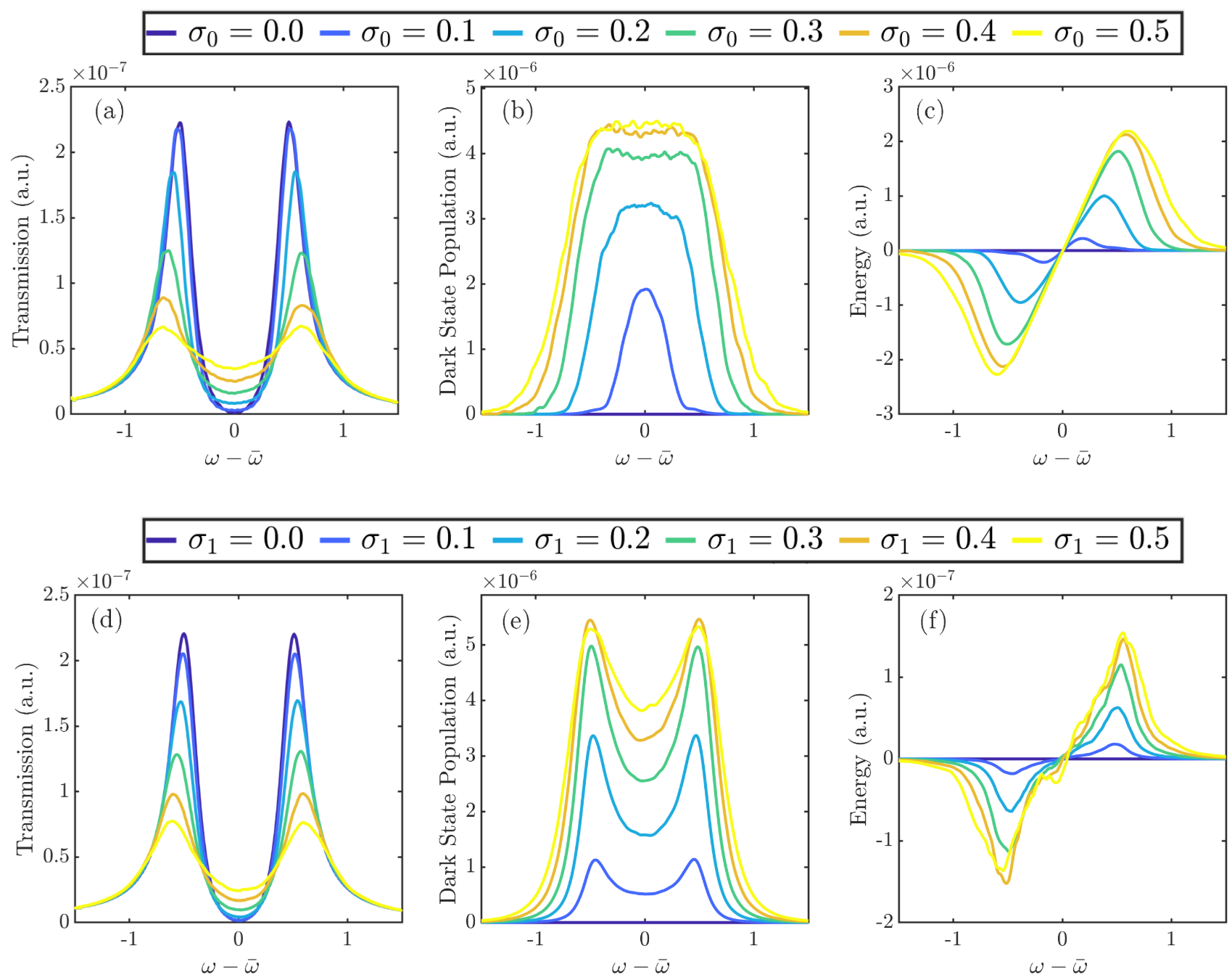}
    \caption{Steady-state properties as a function of the driving frequency $\omega$ with increasing disorder strength $\sigma_0$ for static inhomogeneity (a,b,c) and $\sigma_1$ for dynamic fluctuations (d,e,f) with $\tau_c=2~\text{ps}$. Cavity transmission spectra (a,d) are almost identical, showing reduced intensities. The enhancement of the dark state population is centered at $\omega=0$ for static disorder (b), yet shifted to the polariton frequencies $\omega=\pm0.5$ for dynamic disorder (e). The emitter energy is increased for $\omega>0$ and decreased for $\omega<0$. Note that the maximal enhancement occurs at $\omega\approx \sigma_0$ under static disorder whereas the maximal enhancement occurs at the polariton frequencies $\omega\approx\pm0.5$ under dynamic disorder.}
    \label{fig:StaticDynamic}
\end{figure}

\subsection{Cavity Transmission, Dark State Population, and Energy Modulations}

We first focus on understanding the distinct effects of the static inhomogeneity and dynamic fluctuations on cavity transmission, dark state population, and emitter energy change.
In Fig. \ref{fig:StaticDynamic}, we plot the steady-state spectra as a function of the driving frequency for the emitter ensemble inside the cavity. 
We explore increasing static disorder strength $\sigma_0$ (panels a--c) and dynamic disorder strength $\sigma_1$ (panels d--f). 
As shown in panel (a) and (d), static and dynamic disorder have a similar effect on the cavity transmission spectra---both reducing the intensity of the polariton peaks and increasing the intensity near the emitter frequency $\omega=0$. This observation suggests that the cavity transmission spectra cannot serve as a measure to distinguish the effect of static and dynamic disorders.

In panel (b) and (e), we notice that the dark state populations are overall enhanced as $\sigma_0$ and $\sigma_1$ increase, but the enhancement occurs at different frequencies for static and dynamic disorders.
Under static disorder, Fig.~\ref{fig:StaticDynamic}(b) shows a single peak at the emitter frequency ($\omega=0$) with inhomogeneous broadening width $\sigma_0$. This broadening corresponds to the increasing width of the Gaussian distribution of the emitter frequencies. 
On the other hand, in the case of dynamic disorder, increasing the disorder strength $\sigma_1$ results in two peaks of dark state population centered at the polariton frequencies $\omega=\pm0.5$.

\begin{figure}
    \centering
    \includegraphics[width=\linewidth]{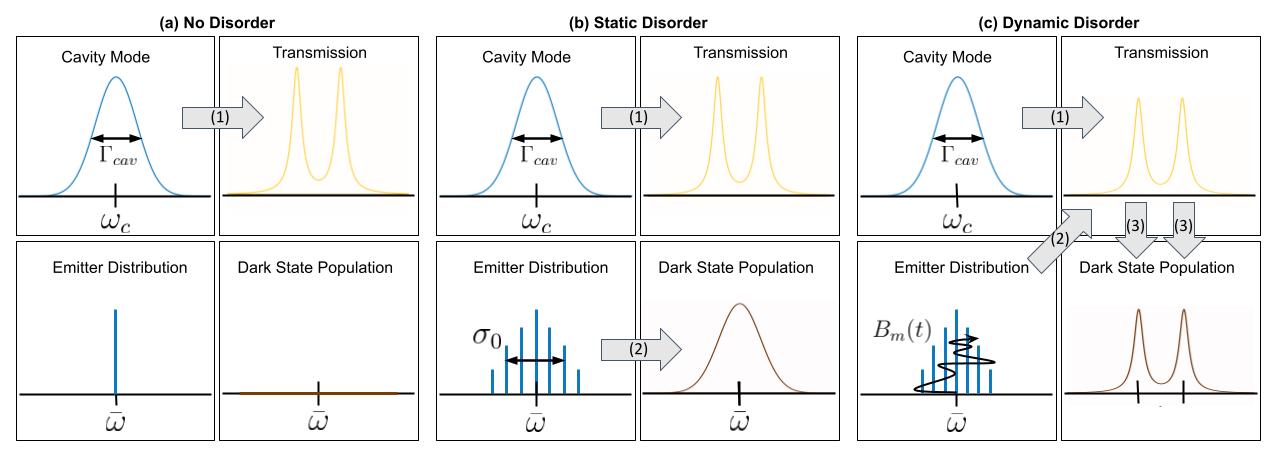}
    \caption{Schematic diagrams for (a) no disorder, (b) static disorder, and (c) dynamic disorder systems. For each panel, the left column shows the cavity mode spectrum (top) and the emitter energy distribution (bottom), and the right column shows the corresponding transmission spectrum (top) and dark state distribution (bottom). (a) Without disorder, the transmission spectrum only has contributions from the polaritons via (1). (b) With static disorder, the emitters nearly resonant with the cavity mode form the polaritons via (1), and the off-resonant emitters populate the dark states via (2). (c) With dynamic disorder, the emitter energy fluctuates with time and becomes resonant with the cavity mode instantaneously, leading to the formation of the polaritons via (2). However, as the instantaneous resonant emitters become off-resonant, they turn into dark states through (3).}
    \label{fig:Schematics}
\end{figure}

To gain more insight, Fig.~\ref{fig:Schematics} illustrates the mechanisms of the polariton formation and the dark states population for a symmetrical system (no disorder) and under static and dynamic disorder. 
Without disorder, the emitters and the cavity photon form the polariton states, and the dark states are not populated, see Fig~\ref{fig:Schematics}(a).
Under static disorder, the emitters that are nearly resonant with the cavity photon form the polariton states while other emitters populate the dark states, which follows the distribution of the emitters as shown as an arrow (2) in Fig.~\ref{fig:Schematics}(b).
As the polariton states undergo mixing with the dark states, the broadening of the dark state distribution as observed in Fig.~\ref{fig:StaticDynamic}(b) leads to suppression of the cavity transmission at the polariton frequencies $\omega=\pm0.5$ in Fig.~\ref{fig:StaticDynamic}(a).
Under dynamic disorder, the emitter energy fluctuates around $\bar{\omega}$ over time and momentarily becomes near-resonant with the cavity photon. In the fast modulation regime, this results in a narrowing of the effective linewidth of the emitter frequency distribution, known as motional narrowing. Consequently, these emitters are more likely to form polaritons rather than remain decoupled from the cavity, as depicted by arrow (2) in Fig.~\ref{fig:Schematics}(c). 
However, the emitters may transition to off-resonant states and populate the dark state, as illustrated by arrow (3) in Fig.~\ref{fig:Schematics}(c). 
When the cavity is driven at the polariton frequency $\omega=\pm0.5$, the cavity transmission is enhanced, followed by an enhancement in dark state population.
Thus, Fig.~\ref{fig:StaticDynamic}(d) and (e) shows an inverse relationship between cavity transmission and dark state population; as the former decreases, the latter increases.
We emphasize that, while the cavity transmission exhibits similar characteristics under both static and dynamic disorder, the underlying mechanisms differ, as shown by their respective dark state population distributions.

Regarding the emitter energy changes, we observe that both disorder effects increase the steady-state emitter energy when the driving frequency is higher than the average molecular energy  ($\omega>\bar{\omega}$) and decrease it when the driving frequency is lower ($\omega<\bar{\omega}$). 
For static disorder (Fig.~\ref{fig:StaticDynamic}(c)), the driving frequency for the maximal change of the emitter energy is correlated with the static disorder strength $\sigma_0$.
For dynamic disorder (Fig.~\ref{fig:StaticDynamic}(f)), however, the driving frequencies for the maximal change of the emitter energy are found at the upper and lower polariton frequencies ($\omega=\pm0.5$), respectively, regardless of disorder strength $\sigma_1$. 
Remarkably, we can understand this feature based on the same mechanism for the dark state populations, suggesting that the emitter energy is dominated by the dark state populations.

\subsection{Transition from fast modulation to static disorder}
\begin{figure}
    \centering
    \includegraphics[width=\linewidth]{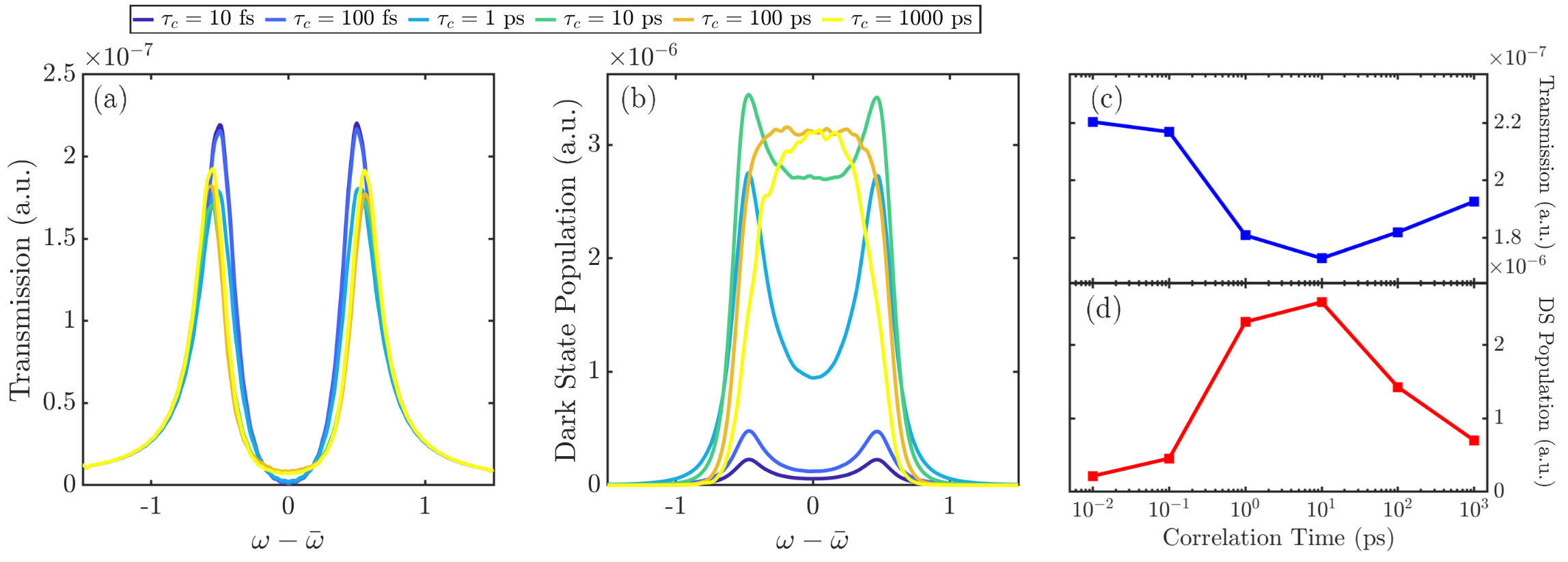}
    \caption{Steady state (a) transmission spectrum and (b) dark state population of the system with a dynamic disorder strength of $\sigma_1=0.2$ at various correlation times. The dark-state population shows the transition from a polariton-like spectrum at short correlation times to a single peak at large correlation times. Scatter plots of the (c) maximum polariton transmission amplitude and (d) its corresponding dark state population amplitude at that frequency. Loss of transmission spectrum amplitudes from the bright state is correlated with a direct increase in the dark state population.}
    \label{fig:Corrtime}
 \end{figure}

With the mechanistic insight into static and dynamic disorder effects, we turn our attention to exploring the transition from fast modulation to static disorder. 
To investigate this transition, we fix $\sigma_0=0$ and $\sigma_1=0.2$ and change the correlation time $\tau_c$.
Fig.~\ref{fig:Corrtime}(a) shows that, as the correlation time $\tau_c$ decreases, the Rabi splitting of the cavity transmission spectrum contracts and the polariton linewidth becomes narrower, which agrees with the previous observation in Ref.~\cite{zhou_interplay_2023}.
The polariton transmission intensity undergoes a transition: it is suppressed when the correlation time decreases from $\tau_c=10^3~\text{ps}$ to $\tau_c=10~\text{ps}$, then increases when $\tau_c<10~\text{ps}$. To understand the cause of this turnover, we calculate the corresponding dark-state population in Fig~\ref{fig:Corrtime}(b). At short correlation times, the dark-state population exhibits two peaks at the polariton frequencies similar to the dynamic disorder results in Fig~\ref{fig:StaticDynamic}(e). As the correlation time increases, the dark-state population transitions to a single peak centered at $\bar{\omega}$ and recovering the static disorder results in Fig.~\ref{fig:StaticDynamic}(b) in the limit $\tau_c=10^3~\text{ps}$. 

To quantitatively analyze the connection between the transmission spectra and the dark-state populations, we plot the maximum transmission amplitude for each correlation time in Fig.~\ref{fig:Corrtime}(c) and their corresponding dark state population at that same frequency in Fig.~\ref{fig:Corrtime}(d). We observe that the polariton transmission amplitude and the dark state population are inversely related. Thus, the population lost from the polariton state (bright with respect to the cavity photon) can be seen to transfer to the dark states through the mixing of these states as induced by disorder.

\subsection{Rabi Splitting Inversion}

\begin{figure}
    \centering
    \includegraphics[width=0.5\linewidth]{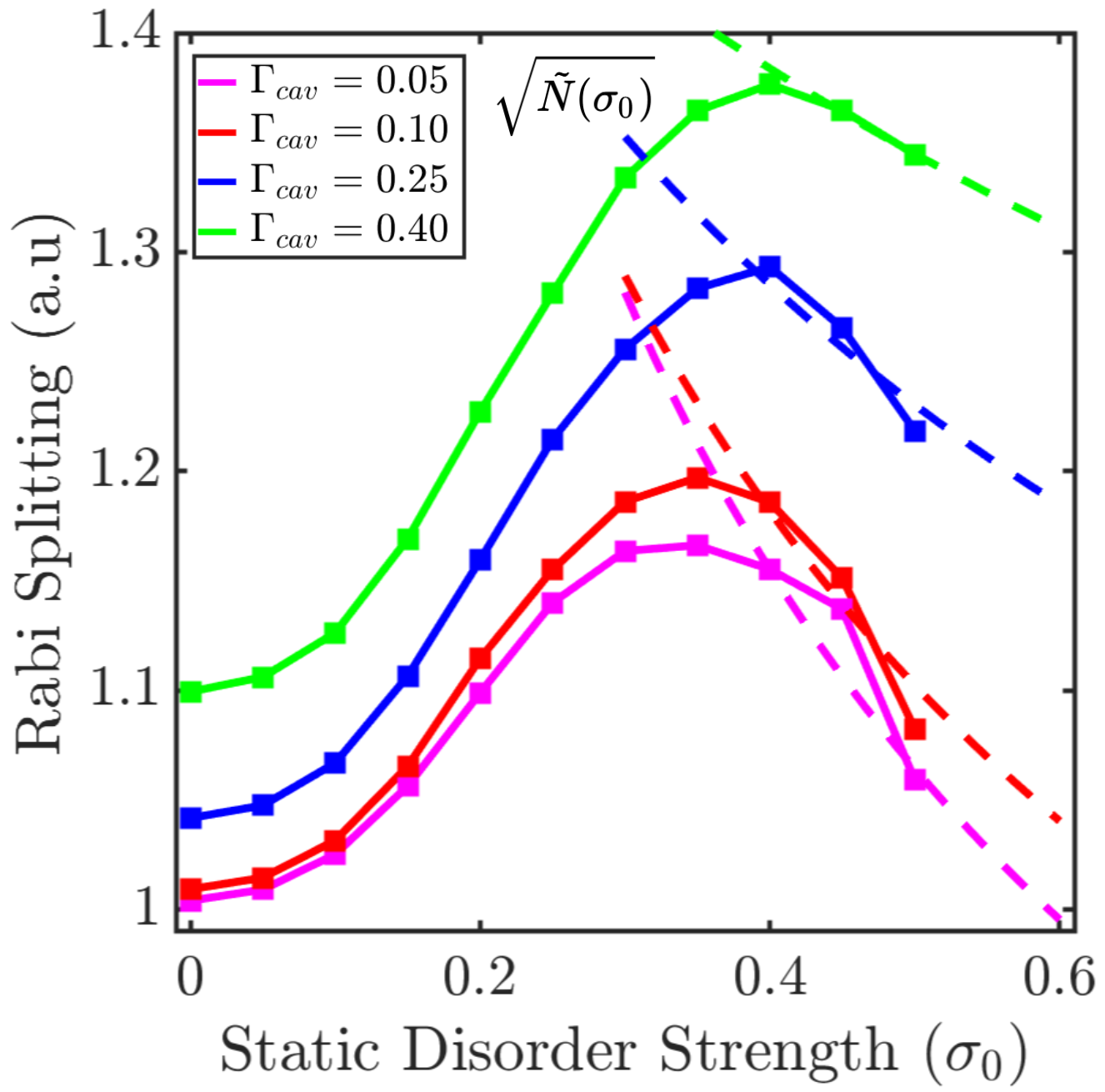}
    \caption{Normalized Rabi Splitting as a function of static disorder strength ($\sigma_0$) for various cavity decay rates ($\Gamma_{cav}$). For small $\sigma_0$, Rabi splitting increases quadratically with $\sigma_0$. The Rabi splitting reaches a maximum and then decreases as $\sigma_0$ increases. We find that the point at which the Rabi splitting is maximized $(\sigma_0^*)$ scales linearly with $\Gamma_{cav}$. The dashed lines are $\sqrt{\tilde{N}(\sigma_0)}$ corresponding to various $\Gamma_\text{cav}$.}
    \label{fig:CavChange}
\end{figure}

With this insight in mind, we now explore the interplay between the cavity photon linewidth ($\Gamma_\text{cav}$) and static disorder strength ($\sigma_0$). 
In Fig.~\ref{fig:CavChange}, we focus on static disorder and quantify the Rabi splitting as a function of $\sigma_0$. 
We observed that the Rabi splitting shows a quadratic scaling with $\sigma_0$ in the small disorder regime, which can be explained by the second-order perturbation theory.\cite{zhou_interplay_2023}
However, as $\sigma_0$ increases, the Rabi splitting reaches a maximum and subsequently decreases. Interestingly, the disorder strength value corresponding to the Rabi splitting maximum ($\sigma_0^*$) scales linearly with $\Gamma_\text{cav}$.
We emphasize that this Rabi splitting inversion is recently observed in spectroscopic experiments with varying the solvent composition. \cite{cohn_vibrational_2022} 

In terms of the mechanisms depicted in Fig.~\ref{fig:Schematics}(b), the Rabi splitting inversion can be understood by considering the overlap between the cavity mode and the emitter distribution. 
At smaller disorder strengths, there is significant overlap between the cavity bandwidth and the emitter distribution, so we can estimate the effective Rabi splitting by treating the disorder as a perturbation, i.e. $\Omega_R=2\sqrt{N}g(1+2\sigma_0^2)$. 
However, at higher disorder strengths, the emitter population is more spread out and overlaps much less with the cavity bandwidth. Consequently, the effective number of emitters coupling to the cavity mode is reduced, leading to the drop in Rabi splitting.
For quantitative analysis, we estimate the Rabi splitting for large $\sigma_0$ by $\Omega_R\rightarrow2\sqrt{\tilde{N}(\sigma_0)}g$ where the effective number of emitters can be approximated by the Voigt profile
\begin{equation}
\tilde{N}=N\frac{\Gamma_\text{cav}}{\sqrt{2\pi^3}\sigma_0}\int_{-\infty}^{\infty}d\omega \frac{e^{-\omega^2/2\sigma_0^2}}{\omega^2+\Gamma_\text{cav}^2}
\end{equation}
Here the Voigt profile comes from a convolution of a Lorentz distribution (cavity photon spectrum) and a Gaussian distribution (emitter ensemble).
In Fig.~\ref{fig:CavChange}, the dashed lines show that the $\sigma_0$ dependence of the Voigt profile captures the decrease of the Rabi splitting for large $\sigma_0$. 

\section{Interplay between Static and Dynamic Disorder}\label{sec:interplay}

Within a solution phase, the fitted FFCF almost always suggests that both static and dynamic disorder are present. Here, we explore the interplay between static and dynamic disorder effects on the Rabi splitting and polariton linewidth. Specifically, we extract the Rabi splitting from the transmission spectra and estimate the linewidth by fitting the spectrum to a sum of two Lorentzian functions. We consider various values of static ($\sigma_0$) and dynamic disorder strength ($\sigma_1$). The Rabi splitting and polariton linewidths are then normalized by the value of the zero disorder case ($\sigma_0=\sigma_1=0$). 

\subsection{Rabi Splitting}

\begin{figure}
    \centering
    \includegraphics[width=0.7\linewidth]{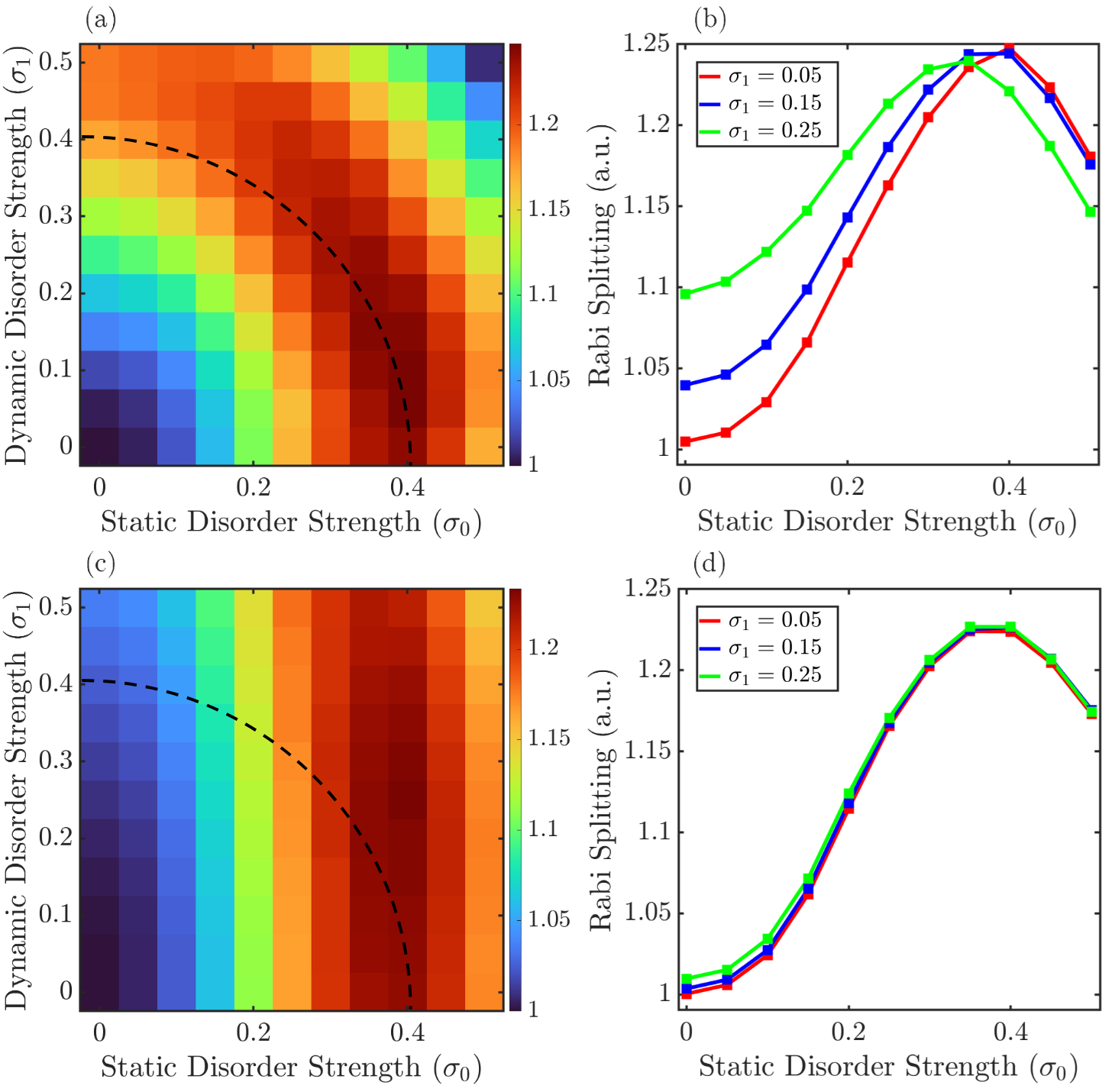}
    \caption{Heatmaps of the normalized Rabi splitting as a function of $\sigma_0$ and $\sigma_1$ for the correlation times $\tau_c=2~\text{ps}$ (a,b) and $\tau_c=0.2~\text{ps}$ (c,d). The quarter-circle dashed line is the Rabi splitting maximum for the case $\tau_c\rightarrow\infty$. 
    For $\tau_c=2~\text{ps}$, the Rabi splitting inversion occurs at lower $\sigma_0$ when $\sigma_1$ increases. The horizontal cuts at $\sigma_1=0.05, 0.15 ,0.25$ in (b) show the Rabi splitting maximum shift to smaller $\sigma_0$. For $\tau_c=0.2~\text{ps}$, dynamic disorder has a smaller effect on the Rabi splitting, and the horizontal cuts are identical.}
    \label{fig:heatmap-Rabi}
\end{figure}

First, in the limit of long correlation time $\tau_c\rightarrow\infty$, the emitter system has effectively two static disorder strengths ($\sigma_0$ and $\sigma_1$), and the effective static disorder strength is $\sqrt{\sigma_0^2+\sigma_1^2}$. In this case, the heatmap becomes symmetric with a quarter-circle shape and the Rabi splitting maximum is depicted as the dashed line in Fig.~\ref{fig:heatmap-Rabi}(a) and (c).

Next, we consider a moderate correlation time $\tau_c=2~\text{ps}$. In Fig.~\ref{fig:heatmap-Rabi}(a), the heatmap of the Rabi splitting notably shows that the maximum occurs at lower $\sigma_0$ as $\sigma_1$ increases while the maximal value of the Rabi splitting remains almost unchanged.
To further emphasize this trend in the Rabi splitting, we plot horizontal cuts at different dynamic disorder values ($\sigma_1=0.05,0.15,0.25$) in Fig.~\ref{fig:heatmap-Rabi}(b), indicating that larger $\sigma_1$ results in a shift of the Rabi splitting maximum. 
This observation suggests that dynamic disorder with moderate $\tau_c$ contributes to the effective static disorder strength $\sqrt{\sigma_0^2+\tilde{\sigma}_1^2}$ where the effective linewidth of dynamic disorder is $\tilde{\sigma}_1<\sigma_1$ due to motional narrowing.

Finally, we consider a shorter correlation time $\tau_c=0.2~\text{ps}$ (i.e, faster dynamic fluctuation) in Fig.~\ref{fig:heatmap-Rabi}(c) and (d).
While the same general trends of the Rabi splitting inversion are observed, the maximum of the Rabi splitting is less influenced by the dynamic disorder, suggesting that fast-fluctuating energy levels do not contribute to the effective static disorder.  
This effect can be attributed to motional narrowing---fast frequency fluctuations lead to reduced effective linewidth of dynamic disorder $\tilde{\sigma}_1$, so that $\sqrt{\sigma_0^2+\tilde{\sigma}_1^2}\approx \sigma_0$.

\subsection{Polariton Linewidth}
\begin{figure}
    \includegraphics[width=0.7\linewidth]{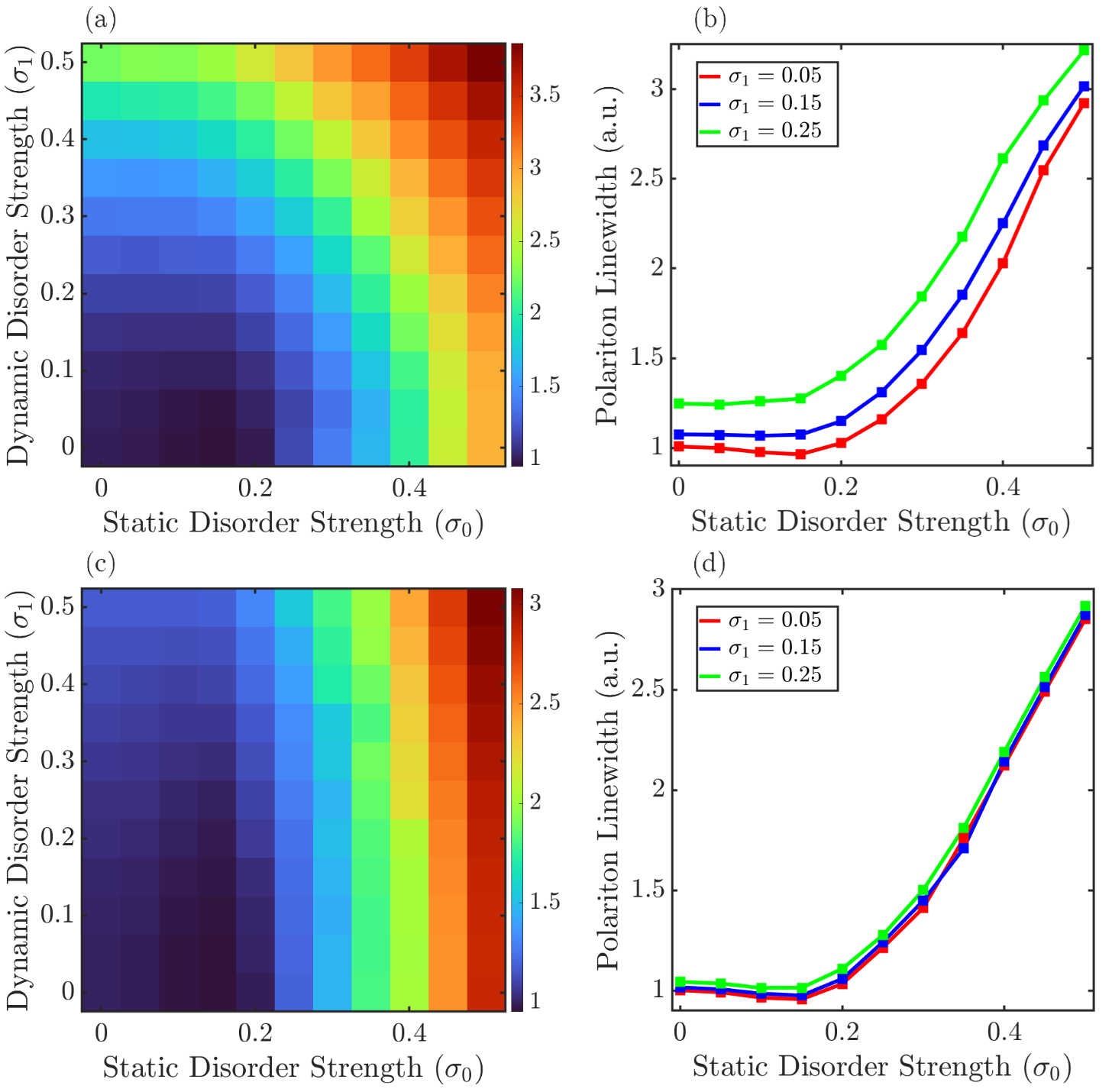}
    \caption{Heatmaps of the normalized polariton linewidth as a function of $\sigma_0$ and $\sigma_1$ for the correlation times $\tau_c=2~\text{ps}$ (a,b) and $\tau_c=0.2~\text{ps}$ (c,d) where (b) and (d) are the normalized linewidth as a function of $a_0$ at various $\sigma_1$.
    When  $\sigma_0>0.2$, the polariton peak is broadened as $\sigma_0$ increases.
    Note that the linewidth in panel (b) is larger than the corresponding linewidth in panel (d) due to motional narrowing. 
    When $\sigma_0<0.2$, the linewidth is narrowed as $\sigma_0$ increases, which is induced by the mixing with the dark states.}
    \label{fig:heatmap-lifetime}
\end{figure}

Regarding the polariton linewidth, the general trend as shown in Fig.~\ref{fig:heatmap-lifetime} is that increasing $\sigma_0$ leads to an overall broadening of the polariton peak in the large $\sigma_0$ regime ($\sigma_0>0.2$). Similar to the Rabi splitting, we can consider a dynamic disorder with a moderate correlation time ($\tau_c=2~\text{ps}$ here) that contributes to an effective static disorder, leading to the simple shift observed in Fig.~\ref{fig:heatmap-lifetime}(b).
Also, a dynamic disorder with a short correlation time ($\tau_c=0.2~\text{ps}$ here) does not influence the polariton lifetime as much (see Fig.~\ref{fig:heatmap-lifetime}(d)). 
Notably, the observed linewidth in panel (b) is significantly larger than the corresponding linewidth in panel (d), which can be attributed to the effect of motional narrowing.

\begin{figure}
    \centering
    \includegraphics[width=0.4\linewidth]{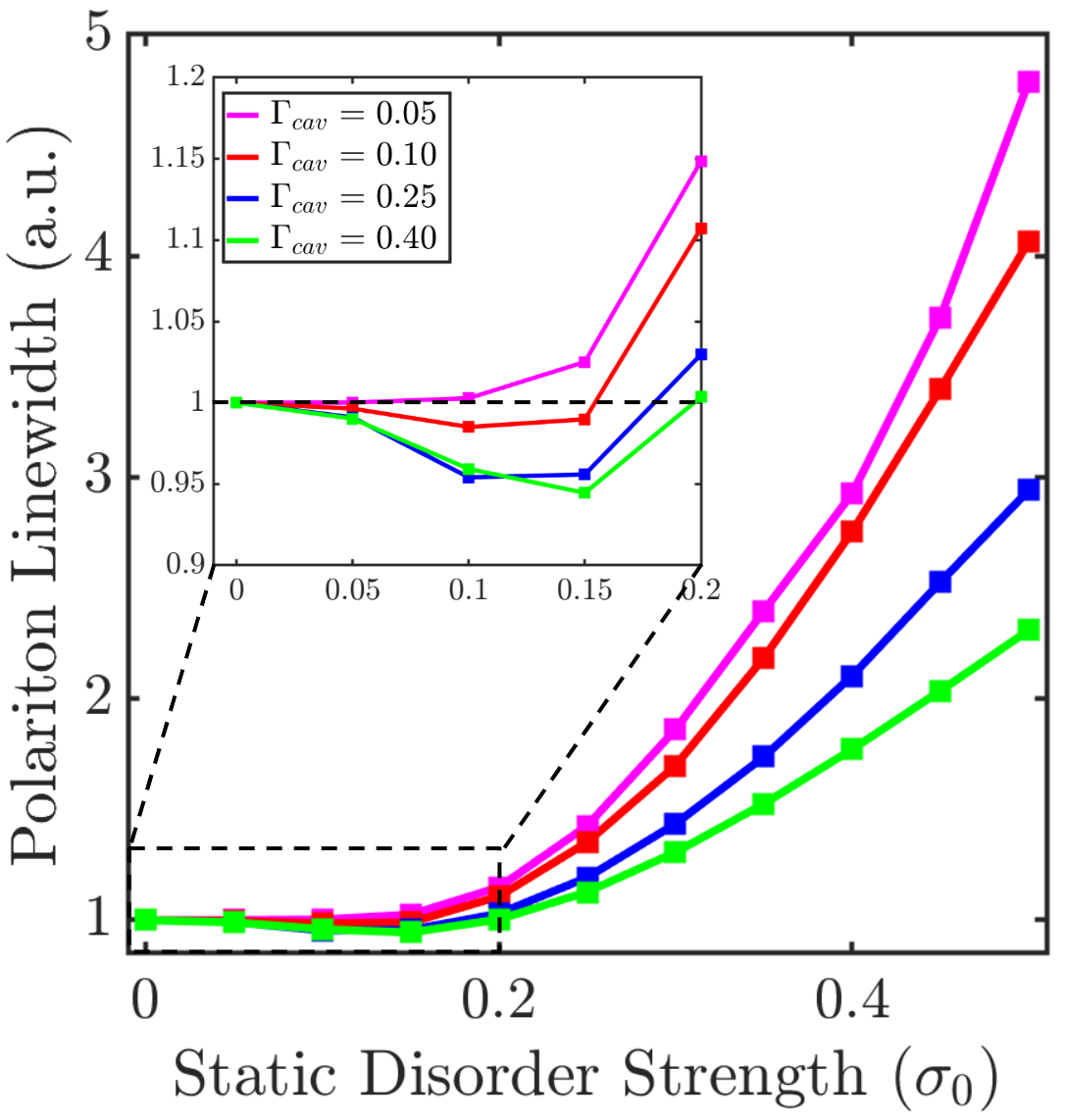}
    \caption{Normalized polariton lifetime as a function of $\sigma_0$ for various $\Gamma_\text{cav}$, fixed $\Gamma_\text{loc}$, and $\sigma_1=0$. For $\sigma_0<0.2$, the polariton linewidth is narrowed at larger values of $\Gamma_\text{cav}$. This narrowing can be understood in terms of the relative contribution from the dark states with long lifetime $1/\Gamma_\text{loc}$ and the cavity lifetime $1/\Gamma_\text{cav}$.}
    \label{fig:lifetime-drop}
\end{figure}

Interestingly, we observe a polariton linewidth narrowing as induced by dark states in the regime $\sigma_0<0.2$ (in comparison to the $\sigma_0=\sigma_1=0$ case). 
To understand this narrowing, we compare the following scenarios.
(\emph{i}) When $\sigma_0=0$, the polariton linewidth can be estimated by $\frac{1}{2}\Gamma_\text{cav}+\frac{1}{2}\Gamma_\text{loc}$, which is dominated by $\Gamma_\text{cav}\gg\Gamma_\text{loc}$ (assuming that the polariton lifetime is additive \cite{deng_exciton-polariton_2010,ying_theory_2024}). 
In our parameter regime, $\Gamma_\text{cav}\gg\Gamma_\text{loc}$ implies that the cavity photon decay dominates. 
(\emph{ii}) When $\sigma_0$ is small, the polariton state is mixed with the dark states that have a longer lifetime ($1/\Gamma_\text{loc}$) than the cavity photon, leading to the drop of the polariton linewidth.
(\emph{iii}) When $\sigma_0$ is large, the spread of the emitter distribution dominates and results in an overall broadening.
Based on these scenarios, we should observe a more pronounced drop when the difference between $\Gamma_\text{cav}$ and $\Gamma_\text{loc}$ increases. 
As shown in Fig.~\ref{fig:lifetime-drop}, this dark state-induced narrowing becomes more pronounced for a larger $\Gamma_\text{cav}$ (low-quality cavity) and a fixed $\Gamma_\text{loc}$.

\section{Conclusion and Discussion}\label{sec:conclusion}
In conclusion, this work offers mechanistic insights into the steady-state optical response of an ensemble of molecular emitters confined within a lossy cavity in the presence of inhomogeneous broadening and frequency fluctuations.
We uncover the contrasting effects of static and dynamic disorder on the dark state population and its correlation with molecular energy change.
Across both static and dynamic disorder, we find a direct influence of distinct dark state populations on cavity transmission spectra; specifically, larger dark state populations correlate with reduced cavity transmission.
This investigation into the combined effects of static and dynamic disorder elucidates intriguing phenomena, including Rabi splitting inversion and dark state-induced polariton linewidth narrowing, as reported in Ref.~\onlinecite{cohn_vibrational_2022}.
Interestingly, we find that the Rabi splitting inversion depends on the combination of static and dynamic disorder and the cavity photon decay rate.
Furthermore, we also rationalize the counterintuitive linewidth narrowing induced by static disorder, attributing it to the contribution of dark states. 
These mechanistic insights highlight the key role of dark states and pave the way for future developments in the fields of polariton chemistry and strong coupling spectroscopy.

Future research should address several factors not accounted for in this study.
First, we assume a single-mode cavity driven by a continuous-wave light source. Future works could extend this study by considering a multimode cavity and by elucidating the transient dynamics induced by pulsed light excitation.\cite{ribeiro_multimode_2022,engelhardt_polariton_2023,mandal_microscopic_2023,tutunnikov_characterization_2024} 
Second, we simplify the system by assuming symmetric emitter-photon coupling, neglecting the potential for intriguing phenomena arising from non-uniform radiative coupling in realistic cavities\cite{zhou_interplay_2023,li_disorder-induced_2024}. As radiative coupling is highly dependent on geometry and electromagnetic environment (including dielectric function and dispersion relations), a future direction of our research will be to incorporate Maxwell-Schr\"odinger equations for multi-emitter systems.\cite{purcell_modeling_2019,perez-sanchez_simulating_2023,sukharev_efficient_2023,clark_harnessing_2024}

\section*{Acknowledgment}
We thank Professor Abraham Nitzan for his pioneering work on strong coupling of molecular ensembles, which inspired this work. We also thank Arnaldo Serrano for the discussion on 2D-IR spectroscopy. 

\section*{Data Availability}
The data that support the findings of this study are available from the corresponding author upon reasonable request. 
\bibliographystyle{unsrt}
\bibliography{references_theta}

\end{document}